**Effect of Laponite® on the structure, thermal stability and barrier properties of**

**nanocomposite gelatin films**

Daniel López-Angulo* (1,4), Ana Mônica Q. B. Bittante (1), Carla G. Luciano (1), German Ayala-Valencia (2), Christian H. C. Flaker (1), Madeleine Djabourov (1, 3), Paulo José do Amaral Sobral (1,4)

(1) Laboratory of Food Technology, Faculty of Animal Science and Food Engineering - FZEA/USP. University of São Paulo, Pirassununga, SP, Brazil, 13635-900.

(2) Department of Chemical and Food Engineering, Federal University of Santa Catarina, Florianópolis, SC, Brazil, 88040-900.

(3) Laboratoire de Physique Thermique, ESPCI-Paris, PSL Research University, Paris Cedex 5, France, 75231.

(4) Food Research Center (FoRC), School of Pharmaceutical Sciences. São Paulo, Brazil, 05508-000.

∗ Corresponding author. Laboratory of Food Technology, Faculty of Animal Science and Food Engineering - FZEA/USP. University of São Paulo, Av. Duque de Caxias norte 225. Jardim Elite. Pirassununga, SP, Brazil, 13635-900. E-mail address: daniel.lopez@usp.br (D. López)

**Abstract**

The effect of Laponite® (a synthetic clay) on the structure, thermal and water vapor barrier properties of nanocomposite gelatin films produced by casting with 0, 4.5 and 15% w Lap/w gelatin, was studied. X-ray diffraction, differential scanning calorimetry, thermogravimetric analysis and Fourier transform infrared spectroscopy measurements were reported. The X-ray diffraction patterns showed dual crystalline structure of the films with collagen-type bundles of triple helices, intercalated inside clay platelets, increasing interlayer distances. Depending on the renaturation of triple-helices and Laponite content, the glass transition temperatures substantially decreased. The amount of helices decreased with Lap concentration, affecting the enthalpy of melting. The nanocomposite gelatin films showed improved thermal stability. Changes of water vapor permeability could be related to the presence of larger free volume of the coils and intercalated structures, facilitating water transfer through the film.

Key words: Laponite®, biodegradable films, biopolymer, glass transition, crystallinity, water vapor permeability.

# 1 Introduction

There is interest in using biodegradable films from renewable resources, such as polysaccharides and proteins. Natural biopolymer-based packaging materials have the potential to extend food shelf life and safety as an innovative packaging and processing technology (Ayala Valencia et al., 2016; Kokoszka et al., 2010; Lee et al., 2018; Li et al., 2015; Staroszczyk et al., 2017). Moreover, the use of biodegradable natural biopolymers reduces the amount of chemical wastes (Avila-Sosa et al., 2010).

Gelatin has been one of the most studied biopolymers due to its film-forming ability and its use as an outer film to protect food from drying and exposure to light and oxygen (Arvanitoyannis, 2002). The ability of some proteins to form networks and induce plasticity and elasticity, are considered beneficial in the preparation of biopolymer-based packaging materials (López et al., 2017).

Although all biodegradable films are not suitable barriers against water vapor and do not have adequate mechanical strength, they can be used as a carrier of antimicrobial substances or preservatives. To overcome the poor mechanical properties of biodegradable films, nanoparticles can be dispersed in the matrix to reinforce it, and to enhance the strength and stiffness of the films (Alexandre et al., 2009). The main reason for this improvement in the mechanical properties of nanocomposite films is the stronger interfacial interaction between the matrix and layered nanoparticles due to the increased exposed surfaces of the nanoparticle layers (Cyras et al., 2008).

The use of clays in food nanocomposite films has been proposed (Avella et al., 2005). Lap (hydrous sodium lithium magnesium silicate) is a synthetic nanoparticle clay, a crystalline layered silicate colloid with structure and composition closely resembling smectic (volcanic) clays. The smectite group includes a variety of clays with the most common ones being montmorillonite and hectorite. Lap nanoparticles have a disk-shape with a thickness of ~1 nm, and a diameter of ~25 nm, much smaller than micron sized natural clays (Perotti et al., 2011). Lap disks have a net negative charge which is balanced by the positive charge of sodium ions. In water, Lap disks hydrate and swell to form clear colloidal dispersions with high stability due to its negative surface charge density of 0.014 $e^-/Å^2$ (Herrera et al., 2004; Nicolai and Cocard, 2000). Lap has been used in agriculture, construction, personal care, surface coatings and the polymer industry (Kumar et al., 2008). It is generally assumed that two factors, the opening up or intercalation of the clay sheets, and the dispersion of the intercalated platelets, determine the properties of dispersions (Rao, 2007).

The objective of this study was to measure the structure and thermal properties of gelatin films containing variable concentrations of Lap. These properties will be correlated to water barrier properties of the nanocomposite films (with Lap) and gelatin films (without Lap).

## 2 Materials and methods

### 2.1. Materials

A pigskin gelatin, type A, bloom 245, average molecular weight $M_w$=5.2 × $10^4$ Da, moisture content=9.98%, (Gelnex South America, Itá, Santa Catarina, Brazil) was used as the biopolymer, and glycerol (Synth, Sao Paulo, Brazil) was used as the plasticizer. Laponite® RD, (Rockwood Additives, Widnes, Cheshire, UK) was used as a synthetic nanoclay. Distilled water was used as solvent.

### 2.2. Film preparation

The nanocomposite films and gelatin films were prepared according to Ayala-Valencia et al. (2016). Briefly, solution A was prepared with 4 g of gelatin/100 g of film-forming solution. Solution B was prepared by dispersing 1% w/w Lap in distilled water at room temperature (20 - 25 ºC). Later, glycerol was added to the Lap dispersion and stirred for 10 min. Then, solutions A and B were mixed to produce film-forming solutions with Lap concentration of 0, 4.5 and 15 g of Lap/100 g of gelatin and a glycerol concentration of 30 g of glycerol/100 g of gelatin.

All the films were prepared by casting; the film-forming solutions were poured into Petri dishes (diameter 14 cm) and dried in an oven with forced air circulation (MA037, Marconi, Sao Paulo, Brazil) at 30°C and controlled relative humidity (55–65%), for 24 h, then the films were cut with a scalpel into rectangular dimensions. Finally, the films were placed over silica gel for 7 days, and then conditioned in desiccators containing saturated solutions of LiCl (11% relative humidity, RH) at 25°C for at least 15 days. Moisture content was determined gravimetrically by using oven drying at 105°C for 24 h (Díaz et al., 2011).

**2.3 X-ray diffraction**

X-ray diffractograms (XRD) of nanocomposite films, gelatin films, and Lap powder were obtained using an X-ray diffractometer (D 5005, AXS Analytical X-ray Systems, Siemens, Berlin, Germany), operating at 40 kV and 40 mA (CuKα radiation λ=1.54056 Å). Samples (3 × 3 cm) were packed in an aluminum frame. The spectra were measured at 25°C between 2θ = 2° and 70°, at 2°/min. From the angular positions (2θ) of the peaks, the corresponding distances ($d$) where calculated using the Bragg relation ($d=\lambda/(2 \sin\theta)$).

## 2.4 Differential scanning calorimetry (DSC)

All films were measured using DSC to determine the glass transition temperature ($T_g$), melting temperature ($T_m$), as well as the enthalpy of melting ($\Delta H$). The analyses were done using a system TA2010 from TA Instruments, Dover, USA, equipped with a cryogenic quench cooling accessory. Samples were placed in aluminum TA pans and weighed with a precision ($\pm 0.01$ mg) balance (PA114C, Ohaus, L.A., USA). An empty pan was used as the reference. These analyses were done under an inert atmosphere (nitrogen gas flow of 45 ml/min), heating samples at 5°C/min from -150 to +150°C.

The $T_g$ was calculated as the inflexion point at the change of the baseline caused by the discontinuity of the specific heat. The $T_m$ was calculated at the maximum of the endothermic peak. The $\Delta H$ value was calculated as the area under the endothermic peak (Sobral and Habitante, 2001), and was expressed in J/g of dry gelatin. Thermal properties were calculated using the software Universal Analysis Version 4.2E Build 4.2.0.38 (TA Instruments).

## 2.5 Thermo gravimetric analysis (TGA)

TGA and derivative thermo gravimetry (DTG) measurements were done using an equipment STA 449 F3 Jupiter® , Netzsch, Berlin, Germany. The measurements were done from 25 to 700°C at a heating rate of 10°C/min and with a nitrogen flow rate of 20 ml/min) to avoid thermo-oxidative reactions (Guo et al., 2013). The samples (~10 mg) were placed in pans with a perforated lid to allow water vapor, and other volatile molecules resulting from the thermal decomposition of the film, to be vented.

## 2.6 Fourier transform infrared spectroscopy (FTIR)

FTIR spectra of dried films were done using a FTIR spectrophotometer (Perkin Elmer Spectrum One, Stuttgart, Germany) from the wavenumber 650 to 4000 cm$^{-1}$ with a resolution of 4 cm$^{-1}$. Sixteen scans were done using the Universal Attenuator Total Reflectance (UATR) accessory. The FTIR spectra were taken in transmittance mode (Ayala-Valencia et al., 2016).

## 2.7 Water vapor permeability (WVP)

WVP was determined gravimetrically, according to the standard ASTM E96 (Ayala-Valencia et al., 2016; Gontard et al., 1992). Briefly, the dried films (size of Petri dishes, disks with diameter of 120 mm) were cut, sized and adjusted to aluminum cells (63.8 mm) containing silica gel (~60 g and 0% RH) and placed in a sealed chamber containing distilled water (100% RH) for a maximum of 7 days. Aluminum cells were weighed (±0.01 g) daily for 7 days to attain steady state permeation. WVP was calculated using Eq. 1:

$$WVP = \frac{\Delta g}{\Delta t} \left( \frac{x}{A \Delta P} \right) \qquad\qquad (1)$$

where $\Delta g / \Delta t$ is the rate of weight change (g/h), x is the sample thickness (mm), A is the permeation (constant) area (0.0032 m$^2$) and $\Delta P$ is the partial pressure difference across the films (3.169 kPa at 25°C) (Ayala-Valencia et al., 2016). The thickness was pleasured using a digital micrometer with 0.001 mm resolution (Mitutoyo, Osaka, Japan), averaging 10 different positions in each nanocomposite film.

## 2.8 Statistical analyses

One way analysis of variance (ANOVA) and Tukey's test of multiple comparisons were done with a significance level of 5% using the Statistical Package for the Social Sciences (SPSS) 10.0 for Windows (SPSS Inc., Chicago, IL, USA).

## 3. Results and discussion

### 3.1 Film-forming solutions (FFS)

The solutions of Lap without gelatin had a high pH (9.5-9.6), while the gelatin solution had a low pH=5.33. The gelatin solutions containing 4.5 and 15% of Lap had pH = 5.97 and 7.03, respectively. All FFS showed some turbidity (Figure 1S).

Turbidity of the FFS showed that strong electrostatic interactions occurred between the two components and the formation of protein-Lap complexes (called coacervates) leading to a liquid-liquid phase separation. Pawar and Bohidar (2010), reported a detailed analysis of the initial phase separation stages of gelatin type A and Lap mixed solutions, and after 10 days observed two distinct phases for the non-gelling solutions. The dense phase is composed mainly of the nanocomposites. The phase separation of nanocomposite films depends on pH.

Several points have been discussed in the literature:

a)    Lap suspensions are stable at pH >10 and show progressive dissolution at pH <9 (Thompson and Butterworth, 1992). However, this effect was found after 7 days of equilibration with agitation. Independently, Pawar and Bohidar (2009), showed that the zeta potential of Lap decreases from -5 to -30 mV with pH varying between 3 and 10. Lap platelets always have negative charges located mainly on the disk faces. Limiting the experimental time may avoid dissolution of Lap at pH <10. Lap is anionic in this pH range.

b)    Gelatin A is a polyampholyte with an isoelectric point around pI ~8-9. The zeta potential is positive at pH <pI and decreases with pH. This protein is cationic below the isoelectric point (pH <8). During the long maturation, gelation starts with the gelatin chains forming triple helices and building a network (Djabourov and Papon, 1983). In the absence of Lap, the number of triple helices does not reach a stable value; the gels evolve with time and temperature.

c)    Upon mixing the two solutions, the pH changes and turbidity appears. Gelatin type A chains bind to Lap platelets by electrostatic interactions modulated by the pH (in the present case, pH ~6-7). Pawar and Bohidar (2009), determined the binding stoichiometry ratio of gelatin type A to Lap and observed the ratio of gelatin A:Lap=1:3, i.e one gelatin chain binds to 3 Lap platelets. Stoichiometric binding corresponds to charge neutralization; most of the positively charged amino acid residues on gelatin type A bind to the negatively charged faces of nanoclay particles. Pawar and Bohidar (2009), did an experiment in dilute gelatin solutions (concentration between $9 \times 10^{-5}$ and $1.56 \times 10^{-2}$ g/cm$^3$).

Gelatin solution concentration was higher, as was Lap concentration, but the electrostatic interactions in these systems should be comparable to the previous reference. The average molecular weight of gelatin was $M_w=5.2 \times 10^4$ Da (López and Sobral, 2016) and of Lap is $M_w=765$ Da (Tawari et al., 2001). Accordingly:

1. In the 4.5% Lap sample, $0.76 \times 10^{-4}$ moles of gelatin are mixed with $2.35 \times 10^{-4}$ moles of Lap which gives a mole ratio gelatin A:Lap ~0.32 close to the ratio 0.33 determined by Pawar and Bohidar (2009) for electrical neutrality. In this case, with 4.5% Lap, the mixed solutions were semi-dilute and the complexes are entangled and cannot sediment rapidly; therefore the solutions become turbid. When they are kept for maturation and drying, they form a gel and form films after equilibration. Gelation prevents sedimentation.

2. In the 15% Lap the mole ratio gelatin A:Lap ~0.1. Lap platelets are in excess with respect to the neutralization ratio. The mixed solutions contain large platelet-rich domains, with some gelatin chains attached to their surfaces and fully intercalated domains. These solutions also undergo gelation, so they do not macroscopically phase separate.

The phase separation between the two Lap populations favors a heterogeneous association between coils and platelets. Suspensions with the two different Lap concentrations formed different structures in the resulting films, shown in particular with the X-ray diffraction measurements.

## 3.2 Films properties

### 3.2.1 X-ray diffraction

The X-ray patterns of gelatin films and of Lap powders (Figure 1a) were measured for comparison with the nanocomposite films (Figure 1b). The gelatin film pattern showed a distinct sharp peak centered on $2\theta=7.2°$ and a broad peak between $10°$ and $50°$ (Figure 1a). Gelatin films were formed after long maturation and dehydration of the solutions in controlled atmosphere, as explained in Section 2.2. In the meantime, gelation occurred due to renaturation of collagen-type triple helices, which grow slowly in solutions. It is a partial renaturation of the native collagen rod, while a substantial part of the gelatin chains remains in a random coil conformation (Djabourov et al., 1988). The triple-helix sequences alternate with the loops and portions of random coils. The triple helices are non-aggregated and randomly oriented in gels. Upon drying in the flat Petri dishes, the gelling solutions become more and more concentrated and the rigid triple helix segments tend to pack into bundles preferably lying flat on the Petri dish surface. The bundles give rise to the sharp peak, observed using X-ray diffraction, whose position depends on the residual humidity in the films. This was shown, in particular by Pezron et al. (1990), that the humidity-sensitive positions of equatorial spots in oriented gelatin fibers are related to the side-by-side distance of the triple helices. The bundles swell when the water content increases: the packing distance between rods (triple helices) increases, with water molecules filling interstitial positions. The large amorphous peak in the diffraction pattern (Figure 1a) is due to solvent (water and glycerol) and random coils of gelatin.

The Lap powder pattern has multiple sharp peaks in the range $0 < 2\theta < 80°$, (Figure 1a, see the red arrows) among which the peak at $2\theta = 5.5°$ is the most important for this study, as it is related to the interlayer spacing of the nanoclay platelets and is sensitive to hydration or intercalation of organic molecules. The basal spacing increases linearly upon water adsorption in Lap powders (the peak position shifts towards smaller angles). The clay swells when water molecules are adsorbed, the interlayer distance increases from 11.5 to 20 Å with $5 < RH < 94\%$ (Valencia et al. (2018). The other intensity peaks of the Lap powder are seen in Figure 1a.

The nanocomposite films contain a superposition of gelatin and Lap patterns, which shows the coexistence of two crystalline organizations mainly identified by the diffraction peaks at small angles ($2\theta < 10°$). The broad amorphous peak, due to gelatin and solvent becomes non-symmetrical in the 15% Lap (number 3, Figure 1b).

To identify the interlayer spacing of Lap, the difference between the intensities diffracted by the nanocomposite films and the gelatin films in the range $2\theta < 10°$ were plotted (Figure 2S). From this figure, it was possible to calculate the positions of the diffraction peaks (Table 1). The corresponding distances, in gelatin films (collagen type bundles of triple helices), in Lap powders and in the nanocomposite films were also calculated (Table 1).

These data allow the following discussions:

The interlayer spacings of Lap in nanocomposite films increased substantially, compared to Lap powder. The interlayer spacings deduced from the diffraction peaks are $d_{interlayer}$=25 and 22 Å with 4.5 and 15% Lap, respectively.

According to Valencia et al. (2018), the interlayer distance in Lap is a minimum of 10-11 Å in dry powders. The gelatin network traps the Lap platelets, and allows them to stack in piles, which was seen on the X-ray patterns of the films. Intercalation of the nanoclay with gelatin led to an interlayer swelling in the 4.5% Lap sample, 25-11 of 14 Å and in the 15% Lap sample, 22-11 of 11 Å. In the initial Lap powder (Figure 1a) the interlayer distance close to $d_{interlayer}$=16 Å, was observed.

The diffraction peak due to the bundles (packing) of triple helices is $d_{collagen}$=11.8 Å in the gelatin film, with a slight increase in the nanocomposite films. Small changes of $d_{collagen}$ suggested local changes of hydration in the films.

The packing distance $d_{collagen} < d_{interlayer}$ is compatible with the intercalation of a layer of bundles between Lap platelets in the nanocomposite films. Electrostatic attraction promotes the adsorption of gelatin coils onto the surface of the Lap platelets in the initial stages.

While gelatin chains change their conformation from coil to triple helix during maturation, the complexes with Lap are maintained including the triple helices between the platelets. The 15% Lap shows a diffraction pattern after maturation that was interpreted as more distinctly phase-separated structures allowing larger stacks of platelets to form, including also the adsorbed gelatin chains; the interlayer was large, 22 Å, and compatible with the packing of triple helices formed during gel formation. Figure 2 shows schematically the dispersion of platelets during the different steps from solution to films and the conformation changes of the gelatin coils.

### 3.2.2 DSC

Regardless of Lap concentration, in the first scan, all the films showed a typical behavior of partially crystalline materials (Figure 3a), only one glass transition ($T_g$) was observed followed by a melting endothermal peak ($T_m$).

The change of glass transition temperatures is a consequence of the interactions with Lap: compared to $T_g$ of gelatin films, in nanocomposite films with 4.5% Lap, $T_g$ decreases by 8°C, and increases again in the 15% Lap films (Table 2). There is a large debate about the possible changes of glass transition temperatures due to confinement of polymers in nanocomposite structures: either a decrease or an increase in $T_g$ have been observed, which depend on the state of the dispersion and the chemical structures of the polymer and of the "walls". For example, in nanocomposite films prepared by dispersion of montmorillonite clay and a miscible blend of polymers (phenylene oxide/polystyrene), Tiwari and Natarajan (2011), observed that the glass transition temperature was not affected by the fraction or the nature of the organoclay. Instead, Nagarajan et al. (2015), observed an increase in $T_g$ value of fish gelatins with a nanoclay Cloisite $Na^+$, crosslinked with an ethanolic extract.

Jackson and McKenna (1991), observed that the glass transition temperature of organic liquids shifted to lower temperatures when they are confined to small pores in glasses with pores down to 4 nm. Varnick et al. (2002), interpreted the changes of $T_g$ of polymer chains using molecular dynamics simulations. They observed by modeling, that strongly attractive walls led to increased $T_g$ because the motion of the chains is slowed with respect to the bulk, near the walls. A weaker attraction has the opposite effect: compared to the bulk, the authors noticed that the molecules relaxed faster. This situation leads to a decrease of $T_g$.

With 4.5% Lap, the Tg decreased. This corresponds to homogeneously distributed polymers and platelets. This effect can be interpreted as a weak attraction between positively charged amino acids of gelatin chains and the negatively charged surface of the platelets. The basal distance of the platelets in the films is large. The free volume of the amorphous coils inside the nanocomposite could be increased, compared to bulk gelatin with the same conditions. As a consequence, the Tg is lowered and the mobility of the amorphous parts is increased (faster relaxation) at the same temperature. The change of $T_g$ is likely to have significant consequences on the permeability of the films at room temperature.

The second scan (Figure 3b) shows typical thermograms of amorphous materials, since the crystal structure was totally melted during the first heating, a single $T_g$ is observed. The $T_g$ calculated in the second scan was lower ($\Delta T_g$=-12.7°C) than $T_g$ of the first scan (Table 2), because crystallinity affects this property (Sobral and Habitante, 2001). The amorphous gelatin segments in the semi-crystalline films have a reduced mobility compared to totally amorphous films. This property may be related to the crosslinking effect of triple helices in the semi-crystalline films which increases $T_g$ by comparison with amorphous, un-crosslinked films, according to Nagarajan et al. (2015).

This behavior was also observed in the Lap nanocomposites. A significant decrease in $T_g$ (second scan) compared to the first scan was observed. In the 4.5% Lap the decrease is $\Delta T_g$=-18 °C and in the 15% Lap, the decrease was -17 °C. $\Delta T_g$ were larger in nanocomposite films compared to gelatin films. The contribution of Lap to $T_g$ depends on the microstructure of the films: although there is only one $T_g$, the spread of the values is important in the second scan of nanocomposite films compared to the second scan of gelatin films (Table 2), which might suggest that gelatin was in different environments.

The melting temperature of gelatin films, $T_m$=96.3°C, was higher than with 4.5% Lap ($T_m$=89°C). The decrease of melting temperature may be a consequence of interactions between gelatin and Lap; the helical structure contains more defects, such as loops or shorter sequences, resulting in less stable structures. At 15% Lap the melting temperature increased back to the initial value, but showed a larger dispersion. As explained in section 3.1, these films showed more segregated structures; gelatin-rich coexist with Lap-rich domains. Triple helices are present in both domains; the average $T_m$ value is closer to the gelatin-rich domains, while the spread of $T_m$ is due to the segregation effects.

The enthalpy of melting was 21.1 J/g gelatin (Table 2). The enthalpy of melting is related to the triple helix denaturation, regardless of the degree of hydration or plasticization of the film. Elharfaoui et al. (2007) showed that the enthalpy of melting was ~55 J/g for triple helices, close to the enthalpy of denaturation of native soluble mammalian collagens. Thus, from the enthalpy of melting measured in the gelatin films, it was deduced that the weight of helices/weight of gelatin was 38%, as reported in Table 2.

In both nanocomposite films with 4.5 and 15% Lap the enthalpy of melting decreases (p <0.05). The enthalpy decreased because the total amount of renaturated triple helices decreased in the presence of the nanoclay. Hellio-Serughetti and Djabourov (2006), established that the rate of helix formation decreased significantly when gelatin chains are cross-linked, reducing the concentration of helices even after a long period of maturation. The triple helices content decreased, respectively, to 32 and 20% (Table 2). Rao (2007), reported similar thermal parameters in nanocomposite films. Lap interfered with the formation of the triple helix of gelatin, resulting in an increase of the amorphous structure of the polymer: electrostatic interactions between Lap platelets and gelatin in solution and further on, intercalation of the gelatin chains inside the Lap crystalline structure hindered the renaturation of the collagen-type triple helices.

The temperature $T_g^{max} \sim 59°C$, was observed for homogeneous semi-crystalline films of gelatin (first scan) or segregated semi-crystalline nanocomposite films containing 15% Lap (first scan).

The temperature $T_g^{min} \sim 33°C$, was observed for interpenetrated, amorphous films with 4.5% Lap (second scan), where adsorption of gelatin on Lap platelets leads to a homogeneous repartition of the constituents in the nanocomposite film, and increases locally the free volume of the gelatin coils. The difference between the two extremes is large, $\Delta T_g = 26°C$.

$T_g$ reflects the permeability of the films. It is above room temperature in these films, but it is very much dependent on the composition, of the triple helix contribution and it may approach room temperature in extreme cases.

### 3.2.3 TGA

TGA and DTG explore the thermal stability of films over a large range of temperatures. There were three steps in the thermal degradation of the films (Figure 4a, Table 3) (Ayala-Valencia et al., 2016). The temperature range for the first step of thermal degradation was ~25-160°C, which corresponds to the loss of adsorbed low molecular compounds (~4%), especially of the bound water. The second step, in the range of ~140-280°C , is mainly related to the degradation of gelatin chains, which represents a weight loss of ~20%. However, the weight loss of this stage may also be related to water molecules in the innermost layers strongly bound to the gelatin chains by hydrogen bonds. Water strongly bound to the protein in gelatin films may be present in the films up to 160-170 °C (Coppola et al., 2012). The third and main stage with greater mass loss (~40%) was seen between ~276-623°C and may be attributed to the decomposition of more thermally stable structures and glycerol loss (Sobral et al., 2001; Tiwari and Natarajan, 2011). Lap concentration had no influence on these steps, it affected the overall mass change, which decreased from 72.6 to 61.1%, with Lap concentration increasing from 0 to 15% (Table 3).

The DTG curves for nanocomposite films (Figure 4b) show a widening of the second step compared to the gelatin film. This could be related to a strengthening of the gelatin structure with Lap and mainly to the interactions identified with FTIR (see section 3.4).

The incorporation of Lap in gelatin films led to a delay in mass loss in the temperature range of gelatin degradation (Figure 4a). The maximum of the derivative of the mass loss versus time shows a significant decrease in the nanocomposite films (Figure 4b, Table 3) mainly in the second and third steps. Zaidi et al. (2010), observed similar behaviors in the thermal analysis of Cloisite-polylactide nanocomposites. The dispersed Lap sheets provide good barrier characteristics to prevent release of degraded gelatin fragments.

In a more fundamental study, Weiss et al. (2018), determined the decomposition products of different amino acids and their respective gas phases under inert atmosphere, in the temperature range between 50 and 320 °C and detected molar masses between 1−199 Da in the vapor phase. They showed that amino acids decompose endothermally, with heats of decomposition between -72 and -151 kJ/mol, at well-defined temperatures, between 185 and 280°C. Thermal decomposition resulted in 4 gases, mainly $H_2O$, less $NH_3$ and hardly any $CO_2$ or $H_2S$. It was shown that above the decomposition temperature of each amino acid, the DSC traces reached the baselines, while the TGA traces showed a large delay of mass loss that continues to decrease over more than 100°C and did not reach the base lines at the highest temperature investigated (320°C). Therefore, a more accurate interpretation of the TGA results of nanocomposite films would require complementary techniques.

**3.2.4 FTIR analysis.**

The width and intensity of FTIR spectral bands as well as the position of peaks are all sensitive to chemical functions changes and conformations of macromolecules (Kong and Yu, 2007). The distinctive bands in the amide region at 1600-1650 cm$^{-1}$ (amide I), triple helix (1635 cm$^{-1}$), 1500-1550 cm$^{-1}$ (amide II) and ~1237 cm$^{-1}$ (amide III) were observed in the spectra of all samples (Figure 5).

The amide I band (~1600 cm$^{-1}$) of gelatin films originates from C=O stretching/hydrogen bonding in combination with COO$^-$, and this peak shifted to higher wavelengths, around 1634 cm$^1$ for nanocomposite films, suggesting that the different concentrations of Lap led to conformational changes of the secondary structure of proteins (Kong and Yu, 2007). This effect is probably due the interactions of silica groups of Lap with carboxylic groups in gelatin that generates hydrogen bonds, and thereby changes the IR absorption of COO$^-$ groups. Amide II bands of films result from the bending vibration of N-H groups and the stretching vibrations of C-N groups at 1548 cm$^{-1}$. Amide III bands are related to vibrations in C-N and N-H groups of bound amide or vibrations of CH$_2$ groups of glycine, and were seen at ~1237 cm$^{-1}$ (Figure 5).

The incorporation of Lap, regardless of the concentration, led to the appearance of bands at ~995 cm$^{-1}$, which can be associated with the C-O vibration of glycerol and Si-O-Si stretching band in Lap (Palkova et al., 2010), as previously reported by Ayala-Valencia et al. (2016). The amide A bands reflecting the stretching vibration of N-H groups occurred at 3288 cm$^{-1}$; in the nanocomposite films, the bands were shifted to higher wavelength. N-H groups of peptides are strongly associated with hydrogen bonds (Kong and Yu, 2007).

### 3.5 WVP

Although not statistically significant, the addition of 4.5 and 15% of Lap in films affected the WVP of nanocomposite films, which slightly increased compared to the gelatin film, which was 0.39±0.02 g mm/m$^2$ h kPa. These changes could be related to the presence of intercalated structures and larger free volume of the coils, facilitating water transfer through the film. Table 4 shows that the thickness of the films containing Lap increases compared to the control film. The film thickness depends on the compaction of the different components during drying. The film is not a pure material, i.e. a solid with a uniform density. Poor compaction facilitates water permeation (Sobral, 2000). The decrease of the compaction in the nanocomposite films is due to several factors: i) larger fraction of amorphous gelatin (data shown in Table 2); compared to the crystalline organization, amorphous phases have larger free volumes (general behavior of solid polymers). ii) voids and cracks of platelets and/or with gelatin may appear during drying. This could be a consequence of the transition from soft to hard films; films are in the glassy state at room temperature (their $T_g$ are above room temperature). iii) Gelatin-Lap films are heterogeneous and locally phase separated. All these factors explain the changes of film compaction.

The results of WVP are consistent with Ayala-Valencia et al. (2016), who reported WVP measurements in gelatin films ($0.37 \pm 0.02$ g mm/m$^2$ h kPa) with no changes upon the incorporation of Lap. The maximum concentration of Lap was 6%.

Li et al. (2015) observed a decrease of the WVP in gelatin-Lap that they attribute to the complex (tortuous) pathway for water diffusion in the presence of the nanoclay. The range of concentration of Lap varied between 5 and 20%. They also observe an increase of the thickness of their films and an increase of the surface roughness. However, they did not do X-ray diffraction measurements. They used a large amount of glycerol in their gelatin stock solutions (5% w/v gelatin and 40% w/w glycerol) and also add 60 mM of NaCl. The composition of the stock solutions probably influences the triple helix renaturation and the presence of salt modifies the interactions between gelatin and Lap.

Farahnaky et al. (2014) reported on gelatin-Cloisite films (Cloisite is a natural montmorillonite modified with a quaternary ammonium salt) that WVP declined with increase of clay content. Again, no X-ray or DSC studies were reported. Their initial value (pure gelatin film) of WVP was 0.86 g mm/kPa m$^2$ h which was high compared with this study and with addition of 18% nanoclay it decreased to 0.42 g mm/kPa m$^2$ h, which is closer to the value observed in this study.

Bae et al. (2009) reported that the addition of Cloisite significantly increased the water vapor barrier properties of fish gelatin-clay composite films (decreases ~75%). However their X-ray diffraction diagram showed poor crystallinity of the clay in the film, which contradicts their analysis.

Improvement of permeation properties could result from the tortuosity (winding paths with Lap stacks) as was suggested in the general review by Duncan (2011). The review reports on polymers in the coil conformation, interacting with clay particles.

The Lap forms distinct intercalated structures with a high degree of crystallinity. The basal distance with intercalated gelatin chains is greatly increased. The stacks are not totally spread over the entire film, as some phase separation takes place in particular for the most concentrated Lap films. The free volume of the gelatin coils in the intercalated nano-clays increases. These changes decrease the permeation barrier of the films. The tortuosity in principle, increases only with the most structured piles. At room temperature, the improvement of the barrier properties related to the tortuosity of Lap stacks is counterbalanced by the increase of chain mobility of gelatin coils. The overall result is a slight increase of permeability, which is inconsistent with the benefits expected using the nanocomposite films. In summary, this investigation highlighted the influence of composition and structure on the barrier properties of the gelatin films containing nanoclays; the main factors were: i) the presence of triple helices, ii) the changes of free volume of gelatin chains, iii) the spatial distribution of the clay stacks throughout the film, iv) the changes of the thickness after drying (compaction).

## 4 Conclusion

The presence of Lap improves the thermal stability of the gelatin matrix due to the reinforcement effect of the clay.

In nanocomposite films, a lower amount of triple helix re-assembled, due to the adsorption with electrostatic interactions of gelatin on Lap platelets.

The intercalated structures of the nanocomposite led to a substantial increase of the basal distance of the platelets. The X-ray diffraction patterns showed a dual crystalline arrangement adaptable to the different Lap compositions. This study showed that the collagen-type bundles of triple helices were inside the crystalline stacks of clay platelets and/or independently in the more heterogeneous structures.

The Tg of the nanocomposite films were strongly influenced, intrinsically, by the renaturation of triple helices and extrinsically, by the presence of Lap. The value of $T_g$ of films compared to room temperature influences the barrier properties. It can limit the benefit provided by the winding paths around of Lap sheets. No significant improvement of barrier properties in nanocomposite films was found in comparison with the control. The specific properties of the nanocomposite films of gelatin and Lap are related to their structure and could be used in new applications.

## Funding

The authors would like to thank the Brazilian National Council for Scientific and Technological Development (CNPq) for the research grant number 150347/2016-2; the São

Paulo Research Foundation (FAPESP) for the grant by Food Research Center (FoRC, 2013/07914-8)-Brazil, and the PhD fellowship number 2012/24047-3; and finally, the Coordination for the Improvement of Higher Education Personnel (CAPES).

**Conflict of interest.**

The authors confirm that they have no conflicts of interest with respect to the study described in this manuscript.

**FIGURE LEGENDS**

Figure 1. a) X-ray diffraction patterns of gelatin films and Laponite powder; b) X-ray diffraction patterns of nanocomposite films.

Figure 2. From solutions to films: A) Laponite dispersion in water; Laponite platelets are negatively charged; B) Gelatin solution at 40°C: the chains are in the coil conformation, the global charge is cationic at pH=6-7; C) Gelatin-Laponite mixed solutions, in the initial state just after mixing: complexes form by electrostatic interactions; D) Gelatin-Laponite mixed gels: after maturation, triple helices form and create a network in the gel; E) 4.5% Laponite nanocomposite films with homogeneous intercalation between platelets and bundles of triple helices; F) 15% Laponite nanocomposite film with a heterogeneous structure.

Figure 3. Differential scanning calorimetry thermograms of nanocomposite films and gelatin films: first scan (a) and second scan (b).

Figure 4. a) Thermo gravimetric analysis scans between 20 and 700°C of gelatin films and nanocomposite films, b) and respective derivative thermo gravimetric scans.

Figure 5. Fourier transform infrared spectroscopy patterns for gelatin and nanocomposite films.

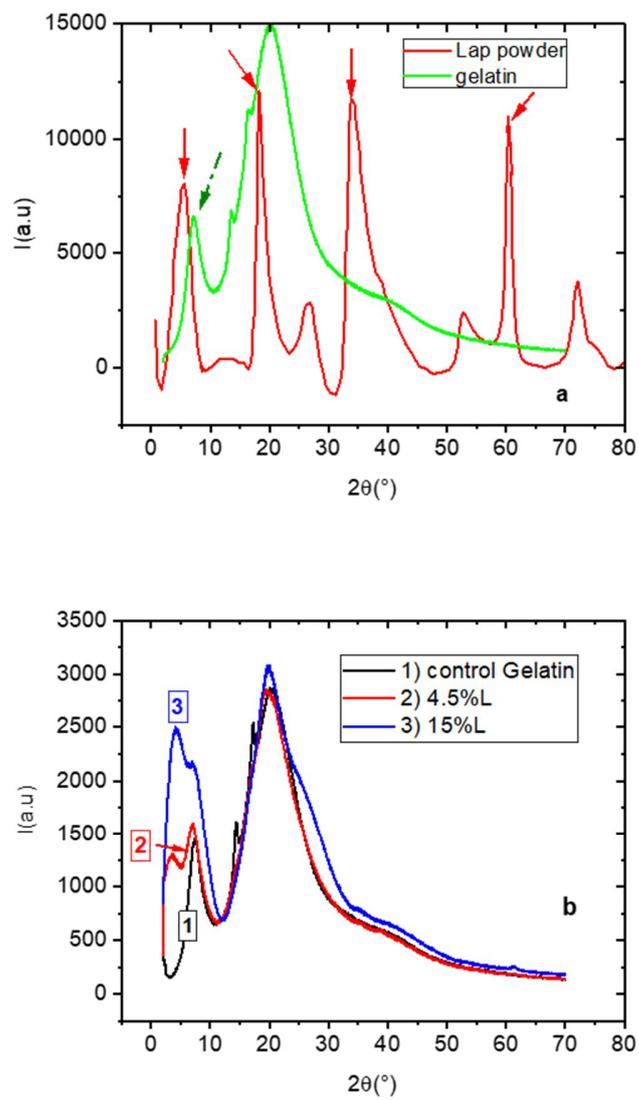

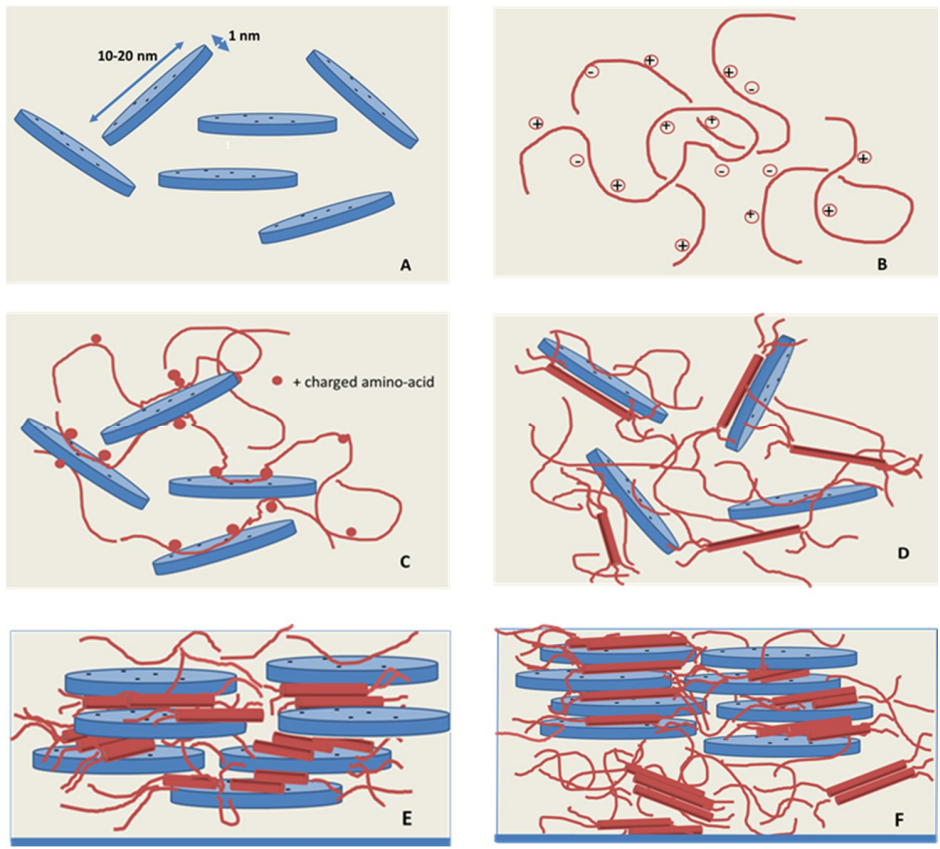

Figure 2

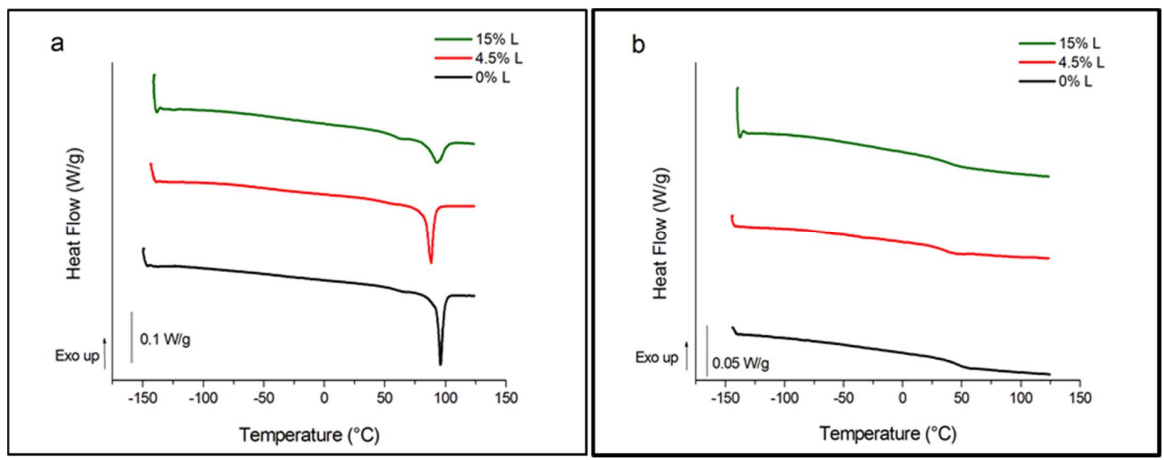

Figure 3

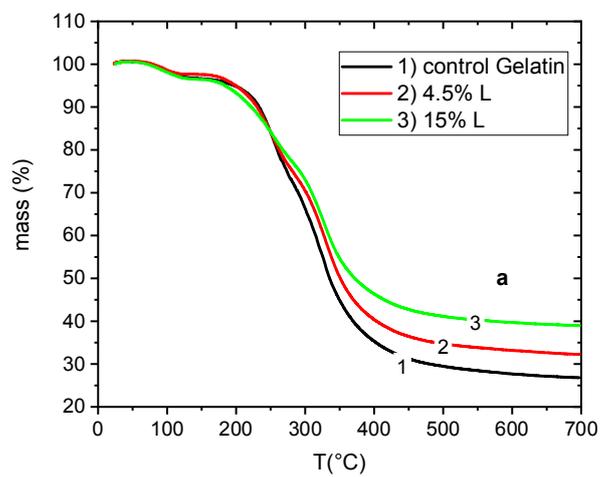

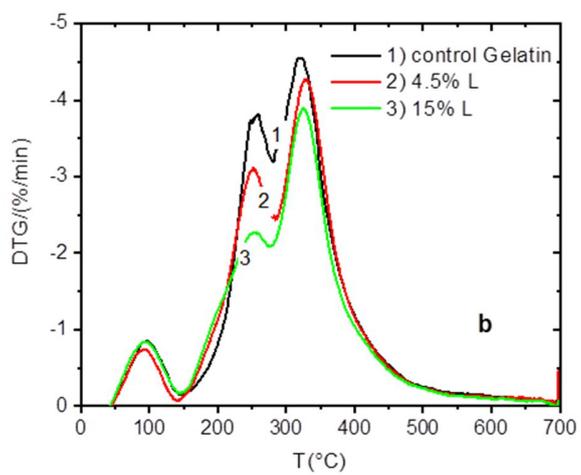

Figure 4

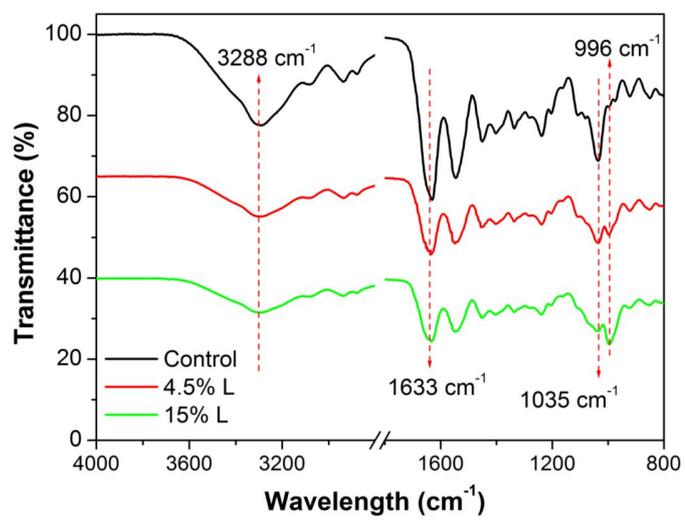

Table 1. Angular positions ($2\theta$) of various peaks and the corresponding distances ($d$) calculated with the Bragg relation in gelatin (collagen type bundles) and Laponite.

| Materials | $2\theta(°)$ | $d$ (Å)* |
|---|---|---|
| Collagen type packing in gelatin films "control", 0 Lap | 7.5 | $d_{collagen}$=11.8 |
| Collagen in nanocomposite films, 4.5% Lap | 7.1 | $d_{collagen}$=12.5 |
| Collagen in nanocomposite films, 15% Lap | 6.9 | $d_{collagen}$=12.8 |
| Lap interlayer in nanocomposite films, 4.5% Lap | 3.5 | $d_{inter}$=25.2 |
| Lap interlayer in nanocomposite films, 15% Lap | 4.0 | $d_{inter}$=22 |
| Lap powder | 5.5 | $d_{inter}$=16 |

Table 2. Thermal parameters obtained by differential scanning calorimetry in nanocomposite films.

| %Lap | First scan | | | | Second scan | |
|---|---|---|---|---|---|---|
| | $T_g$ (°C) | $T_m$ (°C) | Melting enthalpy (J/g of gelatin) | Helices/total gelatin (w/w%) | $T_g$ (°C) | m.c. (%) (d.b.)* |
| **0** | $59.0 \pm 2^a$ | $96.3 \pm 0.4^a$ | $21.1 \pm 0.4^a$ | 38 | $46.3 \pm 0.4^a$ | $14.9 \pm 0.1^a$ |
| **4.5** | $51.0 \pm 1^b$ | $89.0 \pm 0.5^b$ | $18 \pm 1^b$ | 32 | $32.9 \pm 2^b$ | $15.2 \pm 0.1^a$ |
| **15** | $59.5 \pm 2^a$ | $95.6 \pm 2^a$ | $11.0 \pm 1^c$ | 20 | $42.2 \pm 4^a$ | $14.3 \pm 0.1^a$ |

* m.c.: moisture content (dry base). Different letters in the same column indicate significant differences ($p < 0.05$).

Table 3. Thermo gravimetric analysis thermal properties of gelatin films and nanocomposite films.

| %Lap | Temp. range (°C) | $T_{peak}$ (°C) | Mass (%)$_{peak}$ | Derivative (%/min)$_{peak}$ |
|---|---|---|---|---|
| **Water departure** | | | | |
| 0 | 25-160 | 96 | 97 | -0.84 |
| 4.5 | 25-140 | 93 | 96 | -0.74 |
| 15 | 40-145 | 94 | 96 | -0.84 |
| **Gelatin degradation** | | | | |
| 0 | 160-280 | 256 | 81 | -3.78 |
| 4.5 | 140-276 | 254 | 83 | -3.08 |
| 15 | 145-278 | 252 | 82 | -2.27 |
| **Decomposition** | | | | |
| 0 | 280-620 | 320 | 57 | -4.55 |
| 4.5 | 276-618 | 326 | 63 | -4.26 |
| 15 | 278-623 | 327 | 60 | -3.88 |
| **Total mass change (%)** | | | | |
| 0 | -72.6 | | | |
| 4.5 | -67.7 | | | |
| 15 | -61.1 | | | |

Table 4: Water vapor permeability of gelatin and nanocomposite films.

| % Lap | Thickness (μm) | WVP (g mm/m² h kPa) |
|---|---|---|
| 0 | 108±4[a] | 0.39±0.02[a] |
| 4.5 | 118±3[a] | 0.43±0.01[a] |
| 15 | 140±20[a] | 0.49±0.12[a] |

Different letters in the same column indicate significant differences (p <0.05).

**Supplementary Information:**

**FIGURE LEGEND**

Figure 1S. Visual aspect of the different individual and mixed solutions just after mixing: A) gelatin solution; B) aqueous dispersion of 4.5% Laponite; C) aqueous dispersion of 15% Laponite; D) solution of gelatin with 4.5% Laponite; E) solution of gelatin with 15% Laponite.

Figure 2S. Example of subtraction of X-ray pattern of the gelatin films from the patterns of the nanocomposite films in the range $2\theta < 10°$. This procedure allows determining accurately the position of the diffraction peaks associated either to the Laponite stacks or to the bundles of collagen-type triple helices.

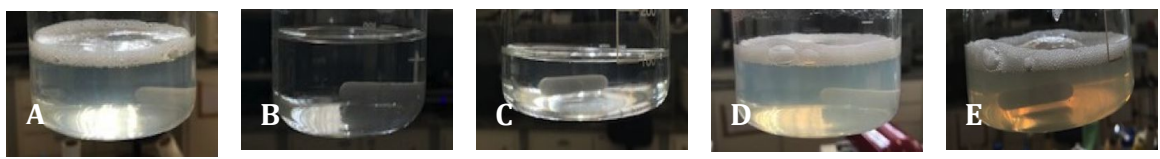

Figure 1S

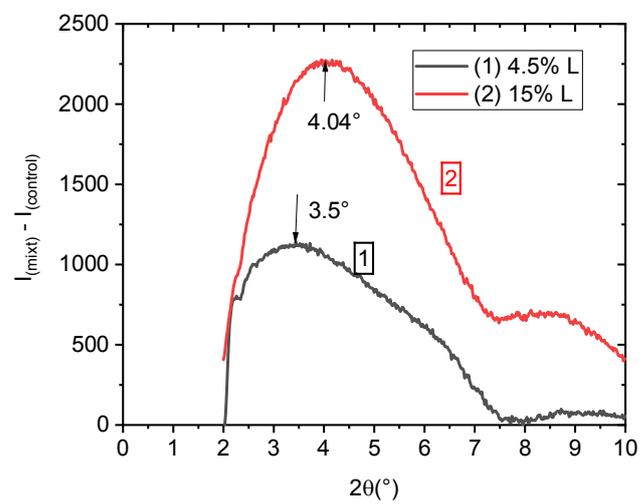

<u>Figure 2S</u>